\documentclass[preprint2]{aastex}

\usepackage{epsfig}

\begin{document}

\title{Nonextensive Effects on Chandrasekhar's Dynamical Friction}

\author{J. M. Silva$^{a,}$\altaffilmark{1}, J. A. S. Lima$^{b,}$\altaffilmark{2} and
R. E. de Souza$^{b,}$\altaffilmark{3}}

\affil{$^a$Centro de Forma\c{c}\~ao de Professores da UFCG, Cajazeiras, PB \\ $^b$Instituto de Astronomia, Geof\'{\i}sica e Ci\^encias
Atmosf\'ericas, USP, 05508-900 S\~ao Paulo, SP, Brasil \\}

\altaffiltext{1}{jmsilva@astro.iag.usp.br}
\altaffiltext{2}{limajas@astro.iag.usp.br}
\altaffiltext{3}{ronaldo@astro.iag.usp.br}

\begin{abstract}
The motion of a point like object of mass $M$ passing through the background potential of massive collisionless particles ($m <<  M$) suffers a steady deceleration named dynamic friction. In his classical work, Chandrasekhar assumed a Maxwellian velocity distribution in the halo and neglected the self gravity of the wake induced by the gravitational focusing of the mass $M$. In this paper, by relaxing the validity of the Maxwellian distribution due to the presence of long range forces, we derive an analytical formula for the dynamic friction in the context of the $q$-nonextensive kinetic theory. In the extensive limiting case ($q = 1$), the classical Gaussian Chandrasekhar result is recovered. As an application, the dynamic friction timescale for Globular Clusters  spiraling to the galactic center is explicitly obtained. Our results suggest that the problem concerning the large timescale as derived by numerical $N$-body simulations or semi-analytical models can be understood as a departure from the standard extensive Maxwellian regime as measured by the Tsallis nonextensive $q$-parameter.
\end{abstract}


\section{Introduction}

It is widely known that a massive object of mass $M$ such as a Globular Cluster passing through a background of non-colliding particles suffers a gravitational effect usually referred to as Dynamical Friction (DF).  Historically, this problem was first studied by Chandrasekhar (1943) who first recognized  its true dissipative nature. He analyzed the idealized case where a point mass moves through an infinity, homogeneous sea of field particles and showed that a fraction of the kinetic energy of the incoming object is transferred to the stellar collisionless population whose distribution was described by a Maxwelllian velocity. 

The DF mechanism is now a classical effect for description and evolution of almost all many-body astrophysical systems. Some examples involve the formation of stellar galactic nuclei via merging of old Globular Clusters (GCs) \citep{Trem1975}, the transformation from non-nucleated dwarf galaxies into nucleated ones \citep{OhL2000}, the behavior of radio galaxies in galaxy clusters \citep{Nath2008},  nonlinear gaseous medium \citep{Kin2009} and the field particles with a mass spectrum \citep{Ciot2010}.  Traditionally, such investigations were carried out in the framework of Newtonian gravity, however, alternative gravity theories like the Modified Newtonian Dynamics (MOND) has also been considered \citep{Nip2008}.

In the last few years, several authors have discussed  the problem related to the dynamical friction timescale ($t_{df}$) of a GC orbiting dwarf galaxies or infalling satellite galaxies in clusters (Read {\it et al.} 2006; Goerdt et al. 2006;  S\'anchez-Salsedo {\it et al.} 2006; Nath 2008; Cowsik {\it et al.} 2009; Inoue 2009; Namouni 2010, Gan {\it et al.}  2010). In particular, when dwarf galaxies have a cored dark matter halo with constant density distribution in its center it has been found  that the DF effects will be considerably modified. Some analyses based on $N$-body simulations \citep{Goerdt2006,Inoue2009} and semi-analytical models have shown that the expected sinking of  CGs to the galactic center may take a time beyond the age of the universe.

In order to explain this suppression effect and the consequent problem of extremely long $t_{df}$, some authors introduced additional ingredients, among them: the loss mass of the  GCs by stripping and heating tidal \citep{Gan2010,Boylan2008}, and the  interaction between the dark matter halo and the GC \citep{OhL2000,Read2006,Inoue2009}.   All these possibilities are based on the standard Chandrasekhar's DF formula derived by assuming  a Maxwellian distribution. Here we investigate a different route by considering that the solution for this problem is related to a proper extension of the underlying statistical approach. 

It is well known that the so-called nonextensive statistical approach  provides an analytical extension of Boltzmann-Gibbs (BG) statistical mechanics which is very convenient when long-range forces are present and/or the system is out (but close) to a thermal equilibrium state. This ensemble theory is based on the formulation of a generalized entropy  proposed by Tsallis (1988,2009)

\begin{equation}
\label{Sq} S_{q} = k_{B}\frac{1-\sum_{i=1}^{W}p^{i}}{1-q},
\end{equation}
which reduces in the limit $q \rightarrow 1$ to the BG entropy $S_{BG} = -k_{B}\sum_{i=1}^{W}p^{i}\ln p_{i}$, since $p_{i}$ is the probability of finding the systems in the microstate $i$, $W$ is the number of microstates and $k_{B}$ is the Boltzmann constant. However, when the index $q \neq 1$, the entropy of the system is nonextensive, i.e, given two subsystems $A$ and $B$, the entropy is no more additive in the sense that $S_{q}(A+B) = S_{q}(A)+S_{q}(B)+(1-q)S_{q}(A)S_{q}(B)$.  The long-range interactions are associated to  the last term on the r.h.s. which accounts for correlations between the subsystems with the index $q$ quantifying the degree of statistical correlations.  Such a statistical description has been successfully applied to many complex physical systems ranging from physics to astrophysics and plasma physics, among them:  the electrostatic plane-wave propagation in a collisionless thermal plasma (Lima, Silva \& Santos 2000), the peculiar velocity function of galaxies clusters \citep{Lavetal98}, gravothermal instability \citep{TaSak2002}, the kinetic concept of Jeans gravitational instability (Lima, Silva \& Santos 2002),  and the radial and projected density profiles for two large classes of isothermal stellar systems \citep{LiSouza2005}. A wide range of physical applications can also be seen in Gell-Mann \& Tsallis 2004 (see also http://tsallis.cat.cbpf.br/biblio.htm for an updated bibliography).

In this letter, by assuming that a self-gravitating collisionless gas is described by the nonextensive kinetic theory (Silva {\it et al.} 1998; Lima {\it et al.} 2001), we  derive a new analytical formula for dynamical friction which generalizes the Chandrasekhar result. As an application, the DF timescale ($t_{df}$) for GCs falling in the galaxies center is  derived for the case of a singular isothermal sphere. This result suggest that the long timescales for GCs can be understood as a departure 
from the extensive regime. In other words, there is no suppression of the DF since the long time can be just the statistical price to pay by the presence of long range forces acting on the gravitational systems. 
   
%The paper is organized as follow: In section 2, we present a review of the basic Chandrasekhar formalism. In  section 3, we deriving an analytic formula for the DF in the context of %the nonextensive kinetic theory. In section 4, we utilize this new kind of the $q$-DF to calculate the dynamical timescale ($t_{fric}$) for a massive objects spirals to the center of %the host galaxy. Finally, in section 5 we presents our conclusions and some prospect for future works.

%\section{The Basic Chandrasekhar Approach}
\section{Dynamical Friction and Nonextensive Effects}

By following Chandrasekhar (1943), the DF deceleration on a test mass $M$ moving with velocity $v_{M}$ in a homogeneous and isotropic distribution of identical 
field particles of mass $m$ and number density $n_{0}$ reads:
\begin{equation}
\label{eq1}
\frac{d{\bf v_{M}}}{dt} = -16 \pi^{2} (\ln \Lambda) G^{2}Mm \frac{\int_{0}^{v_{M}}f(v)v^{2}dv}{v^{3}_{M}} {\bf v_{M}},
\end{equation}
where $G$ is the gravitational constant, $m$ is the mean mass of field stars and $f(v)$ represents their velocity 
distribution. The parameter $\Lambda = p_{max}/p_{min}$ depends on the ratio of the the maximum ($p_{max}$) and minimum ($p_{min}$) impact parameters 
of the encounters contributing to generate the dragging force. 
%In the most of common situations it is considered that $p_{max} \sim L$ being $L$ the size of system and $p_{min} \sim l$ where $l$ is the size of object which can be a GC or black %hole. But in fact a sensitive determination of $\Lambda$ is a delicate problem. 

In the applications of DF, it is usually assumed that the distribution function of the stellar velocity field can be described by a Maxwellian distribution \citep{BT2008,Fellhauer2008}
\begin{equation}
\label{eq2}
f(X_{\star})=\frac{n_{0}}{(2\pi \sigma^{2})^{3/2}}e^{-X_{\star}^{2}},
\end{equation}
where $X_{\star} = v/\sqrt{2}\sigma$ denotes a normalized velocity with $\sigma$ indicating their dispersion.
The integration of (\ref{eq1}) results:
\begin{eqnarray}
\label{eq3} \frac{d{\bf v_{M}}}{dt} = -\frac{4\pi \ln \Lambda G^{2} M \rho(r)} {v^{3}_{M}}H_{1}(X_{M})
{\bf v_{M}},
\end{eqnarray}
where $\rho(r) = n_{0}m$ and the function $H_{1}(X_{M})$ is given by
\begin{eqnarray}
\label{eq3a} H_{1}(X_{M}) = erf(X_{M})-\frac{2X_{M}}{\sqrt{\pi}}e^{-X^{2}_{M}},
\end{eqnarray}
with  $erf(X_{M})$ defining the error function as
\begin{equation}
\label{eq4}erf(X_{M})=\frac{2}{\sqrt{\pi}}\int_{0}^{X_{M}}e^{-X_{\star}^{2}}dX_{\star}.
\end{equation}

%\section{Dynamical Friction and Nonextensive Effects}

Now, in order to investigate the nonextensive effects on the Chandrasekhar theory, let us consider that the stellar field obeys the following power-law (Silva, Plastino \& Lima 1998, Lima, Silva \& Plastino 2001,  Lima \& de Souza 2005): 

\begin{equation}
\label{eq5} f(X_{\star}) = \frac{n_{0}}{(2 \pi \sigma^{2})^{3/2}}A_{q}e_q(X_{\star})
\end{equation}
where the so-called $q$-exponential is defined by
\begin{equation}
e_q(X_{\star}) = \left[1-(1-q)X_{\star}^{2}\right]^{\frac{1}{1-q}},
\end{equation}
and the quantity $A_q$ denotes a normalization constant which depends on the interval of the $q$-parameter. For values of $q < 1$, the positiviness of power argument means that distribution above exhibits a cut-off in the maximal allowed velocities. In this case, all velocities lie on the interval $(0,v_{max})$ and their maximum value is $v_{max} = \sqrt{2}{\sigma} / \sqrt{1-q}$. Taking this into account one may show that the normalization constant $A_q$ can be written in terms of Gamma functions as follows:
\begin{eqnarray}
\label{eq5b} 
%\left
{\begin{array}{ll}
A_{q} = (1-q)^{1/2}(\frac{5-3q}{2})(\frac{3-q}{2})
\frac{\Gamma(\frac{1}{1-q}+\frac{1}{2})}{\Gamma(\frac{1}{1-q})}, \,\,\,\hbox{$q < 1$} 
\\
A_{q} = (q-1)^{3/2} \frac{\Gamma(\frac{1}{q-1})}
{\Gamma(\frac{1}{q-1}-\frac{3}{2})}, \,\,\,\,\, \hbox{$q > 1$}
\end{array}}
%\right\}
\end{eqnarray}

\begin{figure}[ptbh]
\centerline{\epsfysize=45mm\epsffile{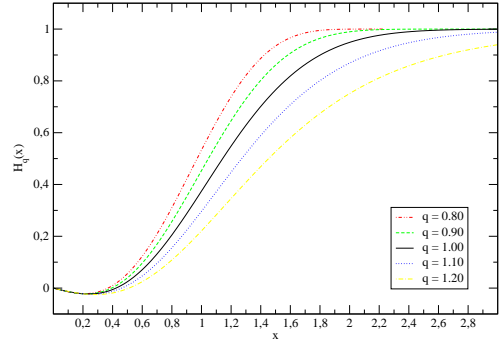}} \caption{The $H_{q}(X)$ function. The solid black curve is the result based on Chandrasekhar theory ($H_{1}(x)$). The remaining  curves show the q-corrections for several values of the $q$-index. } \label{fig1}
\end{figure}

For generic values of $q \neq 1$, the DF (\ref{eq5}) is a power law, whereas for $q = 1$ it reduces to the standard Maxwell-Boltzmann distribution function (\ref{eq2}) since $A_{1} \rightarrow 1$ at this limit. Formally, this result follows directly from the known identity, $\rm{lim}_{d \rightarrow 0} (1 + dy)^{1 \over d} = {\rm{exp}(y)}$ \citep{Abr1972}.  The distribution (\ref{eq5}) is uniquely determined from two simple requirements (Silva {\it et al.} 1998): (i) isotropy of the velocity space, and (ii) a suitable nonextensive generalization of the Maxwell factorizability condition, or equivalently, the assumption that $f(v)\neq f(v_x)f(v_y)f(v_z)$. The kinetic foundations of the above distribution were also investigated in a deeper level through the generalized Boltzmann's equation, in particular, it was also shown that the kinetic version of the Tsallis entropy satisfies an extended $H_q$-theorem (Lima, Silva \& Plastino 2001).

Now, by considering that the power-law distribution (\ref{eq5}) is a valid description for the stellar velocity distribution we conclude that the expression describing the DF in this extended framework takes the following form:
\begin{eqnarray}
\label{eq6} \frac{d{\bf v_{M}}}{dt} &=& -\frac{16\pi^{2}G^{2} (\ln \Lambda) M\rho(r)}{v_{M}^{3}}A_{q} \times \nonumber\\&& \int^{X_{M}}_{0}X^{2}_{\star}e_{q}(X_{\star}) dX_{\star}\bf{v_{M}},
\end{eqnarray}
%The integral above is more easily solved by introducing the following relation
%\begin{eqnarray}
%\label{eq6a}X_{\star}e_{q}(X_{\star})dX_{\star}=-\frac{1}{2}e_{q}^{1-q}(X_{\star})d[e_{q}(X_{\star})].
%\end{eqnarray}
%By performing an integration by parts we obtain:
%\begin{eqnarray}
%&&\int^{X_{M}}_{0}X^{2}_{\star}e_{q}(X_{\star}) dX_{\star} = \nonumber\\&& \frac{1}{5-3q}\left\{\int^{X_{M}}_{0}e_{q}(X_{M})dX_{\star} -  X_{M}e_{q}^{2-q}(X_{M})\right\}. \nonumber\\&&
%\end{eqnarray}
which after an elementary integration can be rewritten as:
\begin{eqnarray}
\label{eq6b} \frac{d{\bf v_{M}}}{dt} = -\frac{4\pi G^{2}\ln \Lambda \rho(r)M}{v_{M}^{3}}\left(\frac{2}{5-3q}\right)H_{q}(X_{M})
\bf{v_{M}}, \nonumber\\&&
\end{eqnarray}
where $H_{q}(X_{M})$ is the general function depending on the $q$-parameter (compare with Eq. (5))

\begin{eqnarray}
\label{eq6c}H_{q}(X_{M}) = I_{q}(X_{M})- \frac{2X_{M}}{\sqrt{\pi}}A_{q}e_{q}^{2-q}(X_{M}).
\end{eqnarray}
In the above expression, the integral 
\begin{eqnarray}
\label{eq6d}I_{q}(X_{M}) = \frac{2A_{q}}{\sqrt{\pi}}\int_{0}^{X_{M}}e_{q}(X_{\star})dX_{\star},
\end{eqnarray}
is the $q$-generalization of the error function (see Eq.(6)).

As one may check, the nonextensive expression for the DF (including the auxiliary functions $H_q$ and $I_q$) reduces to the Chandrasekhar result in the Gaussian limit ($q \rightarrow 1$). It shows clearly that the  collective effect from gravitational interactions of $M$ (with all stars of the field) is strongly dependent on the statistical model.  An interesting aspect of the above formulae is that the results are expressed  by  analytical expressions. In principle, they can be useful for semi-analytical implementations because the easy comparison with the standard approach (see next section). Naturally, we are also advocating here that the idealized framework based on the Maxwellian distribution (Chandrasekhar 1943) may be in the root of some theoretical difficulties shown by $N$-body simulations, like the ones related to the decay orbits of GCs.

\section{Decay of Globular Orbits}

In order to illustrate some consequences of the above derivation, let us now analyze the  nonextensive solution for the decaying orbit of a GC in the stellar galactic field. 
%This is an instructive problem first because it constitutes an intensively studied subject, and, also because the basic results can be expressed by analytical expressions. 
As a GC orbits through the galaxy field, it is subject to DF due to its interaction with the stellar distribution.  By assuming spherically symmetric star distribution, the dragging force  decelerates the cluster motion which loses energy thereby spiraling toward the galaxy center. Therefore, whether the GC is initially on a circular orbit of radius $r_{i}$, it is convenient to define an average DF timescale, $t_{df}$, as the time required for the cluster reach the galaxy center. For the sake of simplicity, we also consider that the mass density distribution of the galaxy is described by the singular isothermal sphere 
\begin{equation}
\label{eq7}\rho(r)=\frac{1}{4\pi G}\left(\frac{v_{c}}{r}\right)^{2},
\end{equation}
with $v_{c}$ being circular speed and $\sigma = v_{c}/\sqrt{2}$ the velocity dispersion. This simplified mass distribution has the benefit of having a planar rotation curve and therefore might be considered as a crude but minimally realistic distribution for the external region of normal galaxies.
Following standard lines, the frictional force undergone by the cluster of mass $M$ moving with speed $v_{c}$ through the stellar field now reads:
\begin{eqnarray}
\label{eq8}F = -\left(\frac{2}{5-3q}\right) G \ln\Lambda \left(\frac{M}{r}\right)^{2} H_{q}(1),
\end{eqnarray}
where $H_{q}(1)$ is the general function (\ref{eq6c}) written in the coordinate $X = (v_{c}/\sigma \sqrt{2}) = 1$. Note also that the integral $I_{q}(X_{M})$ defined in (\ref{eq6d}) now reduces to
\begin{eqnarray}
I_{q}(1) = \frac{2A_{q}}{\sqrt{\pi}} {_{2}F_{1}} \left(\frac{1}{q-1}, \frac{1}{2}; \frac{3}{2}; 1-q \right),
\end{eqnarray}
where ${ _{2}F_{1}}(a, b; c; z)$ is the Gauss hypergeometric function. Either from the  above representation or from the  integral form (\ref{eq6d}), we see that the error function $erf(1)$ is obtained as a particular case in the extensive regime, that is,  $I_{1}(1) = erf(1) \approx 0.8427$ \citep{Abr1972}. It means that $H_{1}(1) = erf(1)-(2/\sqrt{\pi})e^{-1} \approx 0.428$ \citep{BT2008} .

Now, returning to expression (\ref{eq8}), we recall that the dragging force is tangential to the cluster orbits, and, therefore, the cluster  gradually loses angular moment per unit mass $L$ at a rate $dL/dt = Fr/M$. Since $L = rv_{c}$ we can rewritten equation (\ref{eq8}) as
\begin{eqnarray}
\label{eq8b}r\frac{dr}{dt}=-\left(\frac{2}{5-3q}\right) \left(\frac{GM}{v_{c}}\right)\ln\Lambda H_{q}(1).
\end{eqnarray}
By solving this differential equation subjected to the initial condition, $r(0) = r_{i}$, we find that the cluster reaches the galaxy center after a time
\begin{eqnarray}
\label{eq8c}t^{(q)}_{df} &=& \left(\frac{5-3q}{2}\right)\frac{0.5v_{c}r_{i}^{2}}{GM \ln\Lambda H_{q}(1)}.
\end{eqnarray}
This nonextensive timescale for decaying orbits of GCs generalizes  the Chandrasekhar result (see Binney \& Tremaine 2008) which is readily recovered in the Gaussian extensive limit ($q=1$). 

\begin{figure}[ptbh]
\centerline{\epsfysize=45mm\epsffile{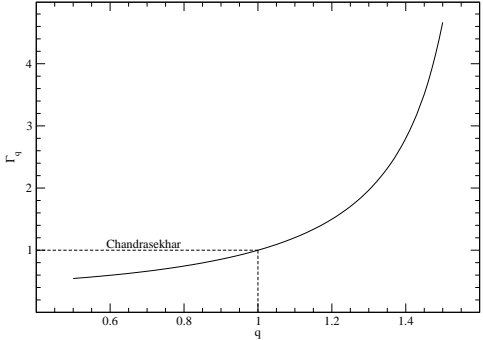}} \caption{Behavior of the relative time scale ratio  $\Gamma_{q}$. We see that for $ q > 1$, the characteristic nonextensive time scale for dynamic friction can be much greater than in the standard Chandrasekhar approach.} \label{fig1}
\end{figure}

At this point, it is interesting to compare the above nonextensive prediction with the standard result based on the Chandrasekhar approach. To begin with, let us  assume typical values for the parameters $r_{i}$, $v_{c}$ and $M$, namely: $r_{i} = 2Kpc$, $v_{c} = 250kms^{-1}$ and $M = 10^{6}M_{\odot}$.  With these choices we get:
\begin{eqnarray}
t^{(q)}_{df} & \approx & \frac{1.14 \times 10^{11}}{H_{q}(1) \ln\Lambda} \left(\frac{5-3q}{2}\right)\left(\frac{r_{i}}{2kpc}\right)^{2} \nonumber\\&& \times \left(\frac{v_{c}}{250kms^{-1}}\right) \left(\frac{10^{6}M_{\odot}}{M}\right)yr,
\end{eqnarray}
which reduces to the standard value in limiting case ($q = 1$) as given by Binney \& Tremaine (2008). The nonextensive 
corrections are more directly  quantified by introducing the dynamic time ratio, $\Gamma(q) \equiv t^{(q)}_{df}/t^{(1)}_{df}$, where  $t^{(1)}_{df}$ denotes the Chandrasekhar result. By using (18) we find
\begin{eqnarray}
\label{eq8d}\Gamma(q) = \left(\frac{5-3q}{2}\right)\frac{H_{1}(1)}{H_{q}(1)}=\left(\frac{5-3q}{2}\right)\frac{0.428}{H_{q}(1)}.
\end{eqnarray}
where the function $H_q(X)$ was defined by Eq. (11). 

In Figure 1, we display the nonextensive corrections for a large range of the nonextensive $q$-parameter. As a general result, we see that the $\Gamma(q)$ ratio is strongly dependent on the q-parameter. The nonextensive time scale is greater or less than the extensive Chandrashekhar result depending on the interval of the $q$-parameter. Note also that $t^{(q)}_{df}$ is greater or smaller than $t^{(1)}_{df}$ if $q > 1$ or $q < 1$, respectively. 

%\subsection{The instantaneous orbital decay}
%It is also widely known that the orbital velocity $v_{c}$ is a function of $r$. Therefore, it is natural to ask about the instantaneous orbital decaying process.
%In order to answer this question we consider  that the angular momentum per unit mass is now given by $L = rv_c(r)$. In this case, Eq. (16) take the following form:  
%\begin{eqnarray}
%\label{eq9}\frac{dr}{dt} = -\frac{4\pi G^{2}M\ln \Lambda }{v_{c}^{2}\frac{d}{dr}[rv_{c}(r)]}\left(\frac{2}{5-3q}\right) r \rho(r)H_{q}(X_{c}(r)),\nonumber\\&&
%\end{eqnarray}
%where $X_{c}(r) = v_{c}(r)/\sigma(r)\sqrt{2}$ and the q-function $H_{q}(X_{c}(r))$ which depends on $v_{c}(r)$ and $\sigma(r)$,  is  defined by% 

%\begin{eqnarray}
%\label{eq10}H_{q}(X_{c}) = I_{q}(X_{c})-\frac{2X_{c}(r)}{\sqrt{\pi}}A_{q}e_{q}(X_{c}(r)).
%\end{eqnarray} 

\section{Conclusions}

We have derived the $q$-dynamic friction force for a point mass moving through a homogeneous background in the context of the nonextensive kinetic theory. Simple and analytical forms were obtained, and, as should be expected, they smoothly reduce to the standard Chandrashekar results in the extensive limiting case ($q = 1$). However, for $q \neq 1$ a large variety of qualitatively different behaviors are predicted when the free parameter $q$ is continuously varied (see Figs. 1 e 2).  As an application, we have discussed the dynamical timescale for a globular cluster collapsing to the center of a massive dark matter halo described by an isothermal sphere. The results presented here suggest that the problem related to the large timescale shown by numerical $N$-body simulations and semi-analytical models may  naturally be solved (with no ad hoc mechanism) by taking a proper $q$-nonextensive distribution with parameter greater than unity. Applications to more realistic density profiles like the lowered nonextensive halos distribution (Silva, de Souza \& Lima 2009; Cardone, Leubner \& Del Popolo) and a detailed comparison with semi-analytical models will be discussed in a forthcoming communication. 
\vspace{0.2cm}

\noindent {\bf Acknowledgments:} JMS is supported by FAPESP Agency and JASL by FAPESP and CNPq (Brazilian Research Agencies).


\begin{thebibliography}{}
\bibitem[Abramowitz \& Stegun 1972]{Abr1972} Abramowitz M. \& Stegun I.A. 1972, Handbook of Mathematical Functions, Dover, NY
\bibitem[Binney \& Tremaine 2008]{BT2008} Binney J. \& Tremaine S. 2008, {\it Galactic Dynamics}. Princeton Univ. Press, Princeton, NJ
\bibitem[Boylan-Kolchin {\it et al.} 2008]{Boylan2008} Boylan-Kolchin M. {\it et al.} 2008, MNRAS {\bf 383}, 93
\bibitem[Cardone, Leubner \& Del Popolo]{Card2011} Cardone V. F., Leubner, M. P. \& Del Popolo A. 2011, MNRAS {\bf 414}, 2265
\bibitem[Ciotti 2010]{Ciot2010} Ciotti L. 2010, AIP Conference Proceedings {\bf 1242}, 117
\bibitem[Cowsik {\it et al.} 2009]{Cow2009} Cowsik R. {\it et al.} 2009, ApJ {\bf 699}, 1389
\bibitem[Chandrasekhar 1943]{Chandra1943}Chandrasekhar S. 1943, ApJ {\bf 97}, 255
\bibitem[Fellhauer 2008]{Fellhauer2008}Fellhauer M. 2008, Lect. Notes Phys. {\bf 760}, 171
\bibitem[Gan {\it et al.} 2010]{Gan2010} Gan J-L. {\it et al.} 2010, Res. Astron. Astrophys. {\bf 10} 1242 
\bibitem[Gel-Mann \& Tsallis 2004]{Gel-Mann2004} Gell-Mann, M., \& Tsallis, C. (ed.) 2004, Nonextensive Entropy:
Interdisciplinary Applications (New York: Oxford Univ. Press)
\bibitem[Goerdt {\it et al.} 2006]{Goerdt2006} Goerdt T. {\it et al.} 2006, MNRAS {\bf 368}, 1073
\bibitem[Gruess {\it et al.} 2009]{Gruess2009} Grues X. M., Lou Y.-K. \& Duschl W. J. 2009, MNRAS {\bf 400}, L52
\bibitem[Hansen {\it et al.} 2005]{Hansen2005}Hansen S. H. {\it et al.} 2005, New Astron. {\bf 10}, 379
\bibitem[Inoue 2009]{Inoue2009} Inoue S. 2009, MNRAS {\bf 397}, 709
\bibitem[Kin \& Kin 2009]{Kin2009}Kin H. \& Kin W.-T. 2009, ApJ {\bf 703}, 1278
\bibitem[Kronberger {\it et al.} 2006]{Kronb2006} Kronberger T. {\it et al.} 2006, A\&A {\bf 453}, 21
\bibitem[Lapenta 2007]{Lap2007} Lapenta, G., Markidis, S., Marocchino, A., \& Kaniadakis, G. 2007, ApJ, 666, 949
\bibitem[Lavagno {\it et al.} 1998]{Lavetal98} Lavagno A. {\it et al.} 1998, Astrop. Lett. Comm. {\bf 35}, 449
\bibitem[Leubner 2005]{Leubner2005} Leubner M. P. 2005, ApJ {\bf 632}, L1
\bibitem[Lima \& de Souza 2005]{LiSouza2005} Lima J. A. S. \& de Souza R. E. 2005, Physica A {\bf 350}, 303, astro-ph/0406404 
\bibitem[Lima, Silva \& Plastino 2001]{Lima2001} Lima J. A. S., Plastino A. R. \& Silva R. 2001, Phys. Rev. Lett. {\bf 86}, 2938, cond-mat/0101030 
\bibitem[Lima, Silva \& Santos 2000]{Lima2002} Lima J. A. S., Silva R. \& Santos J. 2000, Phys. Rev. E {\bf 61}, 3260, 
\bibitem[Lima, Silva \& Santos 2002]{Lima2002} Lima J. A. S., Silva R. \& Santos 2002, A\&A {\bf 396}, 309, astro-ph/0109474 
\bibitem[Namouni 2010]{Nam2010} Namouni F. 2010, MNRAS {\bf 401}, 319
\bibitem[Nath 2008]{Nath2008} Nath B. B. 2008, MNRAS {\bf 387}, L50
\bibitem[Nipoti {\it et al.} 2008]{Nip2008} Nipoti C {\it et al.} 2008, MNRAS {\bf 386}, 2194-2198
\bibitem[Oh \& Lin 2000]{OhL2000} Oh K. S. \& Lin D. N. 2000, ApJ {\bf 543}, 620.
\bibitem[Read {\it et al.} 2006]{Read2006} Read J. I. {\it et al.} 2006, MNRAS {\bf 373}, 1451
\bibitem[S\'anchez-Salcedo {\it et al.} 2006]{SS2006} S\'anchez-Salcedo F. J. {\it et al.} 2006, MNRAS {\bf 370}, 1829
\bibitem[Silva, Lima \& de Souza 2009]{SLS2009} Silva J. M., R. E. de Souza \& J. A. S. Lima 2009, [arXiv:0903.0423].
\bibitem[Silva, Plastino \& Lima 1998]{SPL1998} Silva R., Plastino A. R. \& Lima J. A. S. 1998, Phys. Lett. A {\bf 249}, 401
\bibitem[Silva {\it et al.} 2009]{Sil2009} Silva J. M., R. E. de Souza \& J. A. S. Lima 2009, [arXiv:0903.0423]
\bibitem[Taruya \& Sakagami 2002]{TaSak2002} Taruya A. \& Sakagami M. 2002, Physica A {\bf 307}, 185
\bibitem[Tremaine {\it et al.} 1975]{Trem1975} Tremaine S. {\it et al.} 1975, ApJ {\bf 196}, 407
\bibitem[Tsallis 1988]{Tsallis88} Tsallis C. 1988, J. Stat. Phys. {\bf 52}, 479
\bibitem[Tsallis 2009]{Tsallis2009} Tsallis C., {\it Introduction to Nonextensive Statistical Mechanics: Approaching a Complex World}, Springer (2009)

\end{thebibliography}
\end{document}